\UseRawInputEncoding
\documentclass[reprint, aps]{revtex4-2}
\usepackage{amsmath}
\usepackage{graphicx}
\usepackage{dcolumn}
\usepackage{bm}
\usepackage{float}
\usepackage{appendix}
\usepackage{lettrine}
\usepackage{graphics}
\usepackage{amssymb}
\usepackage{color}
 \usepackage {marvosym}
 \usepackage{ifsym}
 \usepackage{hyperref}
\usepackage{mathrsfs}
\usepackage{caption}
\usepackage{subcaption}
\usepackage{gensymb}
\usepackage{ragged2e}
\definecolor{myred}{rgb}{0.75,0,0}

\begin{document}
\title{Quantum-Enhanced Atomic Sensor via Spin Nonequilibrium Criticality}
\author{Ding Huang$^{1,3}$}
\author{Minwei Shi$^{1,3}$}
\author{Guzhi Bao$^{1,3,4}$}
\email{guzhibao@sjtu.edu.cn}
\author{Keye Zhang$^{2,3}$}
\email{kyzhang@phy.ecnu.edu.cn}
\author{Weiping Zhang$^{1,3,4,5,6}$}
\email{wpz@sjtu.edu.cn}
\affiliation{
$^{1}$ School of Physics and Astronomy, Shanghai Jiao Tong University, Shanghai 200240, China.\\
$^{2}$ Quantum Institute for Light and Atoms, School of Physics, East China Normal University, Shanghai 200062, China.\\
$^{3}$ Shanghai Branch, Hefei National Laboratory, Shanghai 201315, China.\\
$^{4}$ Tsung-Dao Lee institute, Shanghai Jiao Tong University, Shanghai 200240, China.\\
$^{5}$ Collaborative Innovation Center of Extreme Optics, Shanxi University, Taiyuan, Shanxi 030006, China.\\
$^{6}$ Shanghai Research Center for Quantum Sciences, Shanghai 201315, China.}

\begin{abstract}

The sensitivity of quantum sensors is fundamentally constrained by the standard quantum limit (SQL) arising from intrinsic quantum fluctuations. While non-classical resources like squeezing or entanglement can surpass this limit, their utility is often restricted by the extreme fragility of entangled states and the complexity of their preparation. Quantum criticality offers a compelling alternative by harnessing divergent susceptibility to amplify signals without requiring fragile non-classical resources. However, the practical benefit of this approach has remained controversial due to the potential for the simultaneous amplification of quantum noise. Here, we demonstrate a universal protocol for noiseless critical sensing by engineering a light-driven atomic ensemble near a dynamical critical point. Analogous to a Kapitza pendulum near its inverted orientation, the spin system enters a non-equilibrium regime where the signal susceptibility diverges while the quantum noise periodically recedes to its coherent baseline. We exploit this ``noise ebbing'' to create a built-in noiseless amplifier, demonstrating a 3.3 dB metrological gain over the SQL in an atomic magnetometer. Our implementation exhibits intrinsic robustness against common experimental imperfections such as detection losses, establishing non-equilibrium critical dynamics as a practical and versatile paradigm for surpassing the fundamental limits of quantum sensing.
\end{abstract}

\maketitle

\section{Introduction}

The advancement of quantum technologies has spurred a global effort to push the sensitivity of quantum sensors toward the fundamental limits imposed by quantum noise. In atomic ensembles, this limit is defined by spin-projection noise, which set a formidable barrier for high-sensitive measurements across various platforms, including atomic clocks \cite{RN88,RN45,RN89}, inertial sensors \cite{RN86}, and magnetometers \cite{RN90,RN87,RN73,RN74,RN68,RN6} used in fields ranging from biomagnetic imaging \cite{RN91,RN92,RN93,RN94,RN95} and navigation \cite{RN96} to searches for fundamental symmetry violations. While quantum-enhanced strategies such as squeezing and entanglement offer, in principle, a route to circumvent this noise, their practical utility remains constrained by the extreme fragility of non-classical states to decoherence and the prohibitive operational overhead required for their preparation and maintenance \cite{RN75,RN76,RN39,RN77}.

\begin{figure*}[htbp]
	\centering
		\includegraphics[width=1\textwidth]{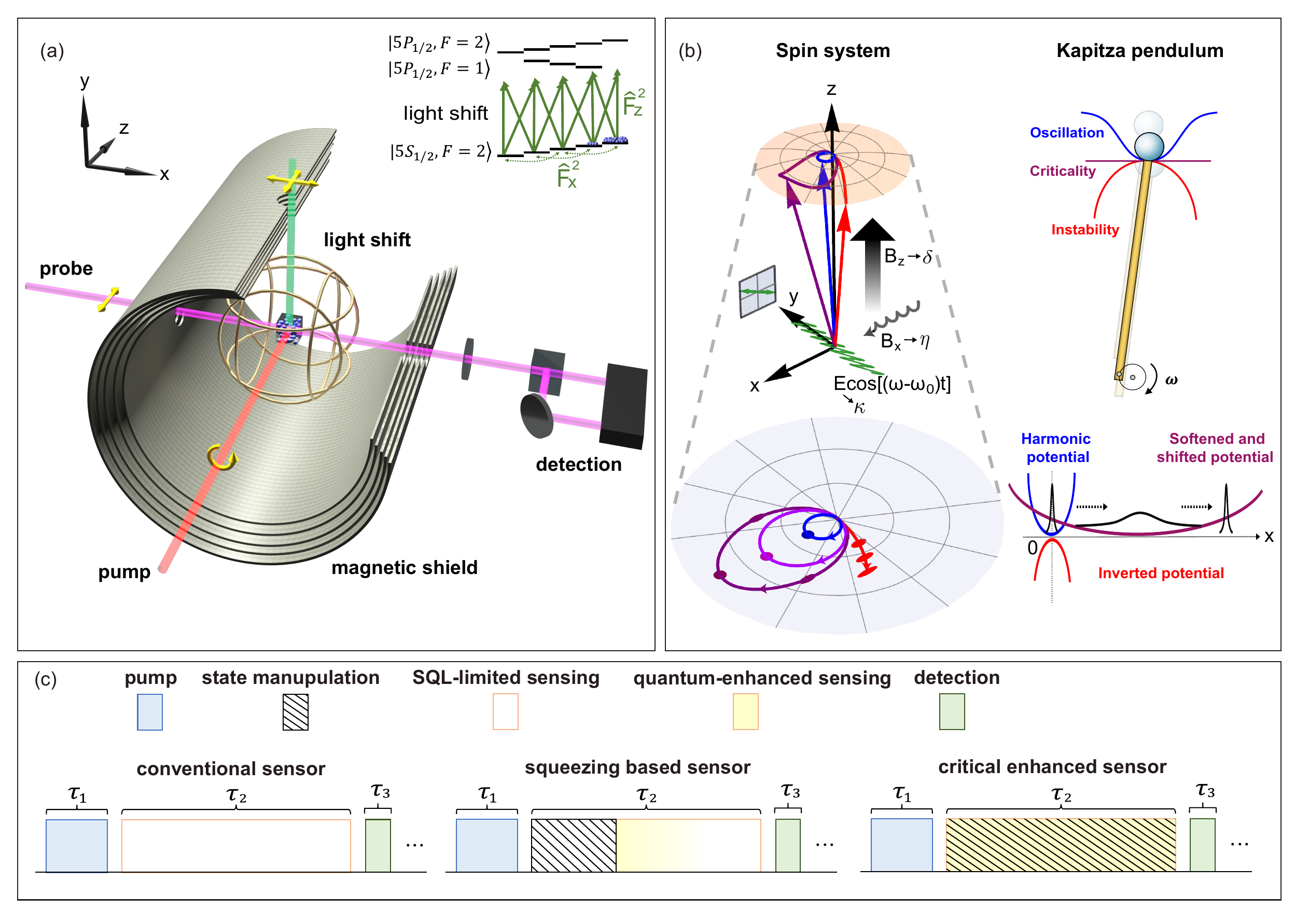}
        \caption{ \justifying \textbf{Experimental setup and operating principles of the criticality-enhanced atomic magnetometer} \\
(a) Experimental setup. A paraffin-coated vapour cell ($30\,\text{mm} \times  30\, \text{mm} \times 30\, \text{mm}$) inside a four-layer magnetic shield. 3D Helmholtz coils provide a $\hat{z}$ bias field and an $\hat{x}$ signal field, with the $\hat{y}$ component actively nulled. Pump, probe, and drive lasers propagate along $\hat{z}$, $\hat{x}$, and $\hat{y}$, respectively. \textit{Inset}: atom-light level diagram.
(b) Dynamical criticality. The left panel shows spin trajectories and fluctuation ellipsoids on the Bloch sphere alongside their equatorial projections where blue curves represent stable harmonic cycles far from the transition. Approaching criticality in purple, the dynamics exhibit expanded orbits and distorted fluctuations before the eventual breakdown of stability beyond the threshold in red. This evolution is mirrored by the inverted pendulum analogy (right up) where the drive amplitude of the fast-oscillating pivot reshapes the effective potential to govern the transition across these distinct dynamical regimes. The corresponding critical evolution of the wave function under the target signal is depicted below (right down).
(c) Sensing sequences. Conventional sensor: polarise ($\tau_1$), precess ($\tau_2$), read out ($\tau_3$). Squeezing-enhanced: part of $\tau_2$ is devoted to squeezing the state.
Criticality-enhanced: spin manipulation and field encoding occur simultaneously during $\tau_2$.}

		\label{fig:1}
\end{figure*}

Critical sensing, characterized by a divergent susceptibility to infinitesimal external perturbations near a phase transition, presents a compelling alternative for enhancing sensor performance without relying on fragile non-classical resources \cite{RN15,RN28,RN20,RN21}. Despite several proof-of-principle demonstrations \cite{Rn16,RN17,RN18,RN49,RN103,RN104}, the question of whether quantum criticality can provide a practical advantage in high-performance sensors remains an open question. These challenges include the amplification of noise alongside the signal \cite{RN79}, the degradation of gain in finite-sized ensembles \cite{RN49}, and the phenomenon of critical slowing down which shifts the sensing bandwidth into low-frequency regimes dominated by technical noise.

A parallel paradigm from classical metrology suggests a way forward. The most sensitive measurements often arise from dynamic rather than static probes. Successful examples include dynamic-mode atomic force microscopy \cite{RN105,RN106} and quartz crystal microbalances \cite{RN108,RN109} which harness the frequency shifts of driven oscillations to detect infinitesimal forces and masses. Unlike static probes that are limited by intrinsic stiffness and suffer from monotonic noise accumulation over time, these dynamic systems exploit non-equilibrium evolution to isolate signals from their environment.

The synergy between dynamic and critical sensing principles establishes a framework that simultaneously eliminates the reliance on non-classical states and overcomes the noise-amplification hurdles intrinsic to quantum critical metrology.
We experimentally implement this protocol in an atomic magnetometer based on a light-driven, room-temperature alkali-metal vapor cell whose dynamics map directly onto the inverted Kapitza pendulum as a paradigmatic model for dynamical criticality. This approach actively exploits a unique asymmetry in the evolution of signal and noise near dynamical criticality. The signal susceptibility diverges while the quantum noise floor periodically returns to its initial coherent‑state level. By engineering this ``noise-ebbing" mechanism, we achieve noiseless signal amplification, providing a robust path to surpass the standard quantum limit (SQL) using only easily prepared coherent states.

Crucially, this criticality-enhanced approach is inherently robust to detection losses and technical noise, which often plague conventional quantum protocols. Synchronizing the measurement with the periodic restoration of spin noise allows us to observe a 3.3 dB metrological gain over the SQL. These results establish non-equilibrium critical dynamics as a practical and versatile resource for quantum-enhanced sensing, offering a scalable route toward sub-femtotesla magnetometry in real-world environments without the need for complex spin squeezing or entanglement preparation.

\section{Kapitza-inspired critical sensing}

While a simple pendulum serves as a paradigmatic model for sensitive sensing in technologies ranging from large‑scale gravitational-wave observatories to microscopic atomic force microscopy, such equilibrium-like systems are typically confined to a static potential minimum. In this regime, sensitivity is fundamentally limited by the standard quantum limit and the inherent trade-off between signal amplification and environmental dissipation.
In such a linear response regime, enhancing sensitivity usually requires a reduction in the restoring force, which simultaneously renders the probe increasingly fragile and susceptible to thermal or technical noise. We transcend this conventional paradigm by operating within a non-equilibrium framework inspired by the Kapitza pendulum \cite{RN62}. Unlike its static counterpart, our protocol employs a high-frequency periodic drive to stabilize an otherwise unstable inverted configuration, ensuring that sensitivity is determined by the precision of experimental control near the stability threshold rather than the intrinsic stiffness of the potential.

We implement this protocol using a collective spin ensemble in an alkali-metal vapor cell where a static magnetic field plays the role of gravity and an off-resonant tensor light shift provides the high-frequency Kapitza drive. While conventional atomic magnetometers rely on Larmor precession away from a polarized orientation, our scheme establishes a fixed critical point through Kapitza stabilization, yielding a susceptibility well above the system's natural response. Near this threshold, the resulting critical dynamics strongly amplify the atomic response to the target magnetic field. This synergy between dynamic and critical sensing enables the magnetometer to achieve a signal-to-noise ratio that surpasses the standard quantum limit, effectively circumventing the sensitivity constraints typical of traditional Larmor precession-based sensors.

Our scheme is based on a room-temperature ensemble of $^{87}\mathrm{Rb}$ atoms. The magnetic sensor consists of a cubic glass cell, placed inside a multi-layer magnetic shield with three-axis Helmholtz coils for active magnetic-field control as illustrated in Fig. \ref{fig:1}a. The atoms are optically pumped into the fully polarized state $\left|5S_{1/2},F=2,m_F=2\right\rangle$ along the $\hat{z}$-axis using two circularly polarized 795 nm lasers as pump and repump beams. These address the transitions $\left|5S_{1/2},F=2\right\rangle \to \left|5P_{1/2},F=2\right\rangle$ and $\left|5S_{1/2},F=1\right\rangle \to \left|5P_{1/2},F=2\right\rangle$ respectively.
A static bias magnetic field is applied along $\hat{z}$-axis, producing a Zeeman splitting with Larmor frequency $\delta = \gamma B_z$, where $\gamma$ is the gyromagnetic ratio. Additionally, an off-resonant light beam is introduced along the $\hat{y}$-axis to generate a periodic perturbation. By setting its polarization at the angle of $35.26^\circ$ relative to the $x$-axis and splitting it into two orthogonal parts, the resulting tensor light shift \cite{RN7} effectively couples electron and nuclear spins, leading to a periodically driven atomic Hamiltonian
\begin{equation}
\hat{H}_1 = \sum_{m,m'}\left( m \delta \hat{\sigma}_{m,m} +  \mu_{m,m'}\boldsymbol{E}\cos{[(\omega-\omega_{0}) t]}\hat{\sigma}_{m,m'} + h.c.\right),
\label{eq1}
\end{equation}
where $\omega_{0}$ is the frequency of the relevant excited states, $\boldsymbol{E} = \vec{E}_{+} + \vec{E}_{-} + \vec{E}_{\pi}$ denotes the amplitude vector of the optical field, $\omega$ its frequency, and $\mu_{m,m'}$ the dipole matrix element for the atomic transition associated with the density operator $\hat{\sigma}_{m, m'}$.
In this picture, the Zeeman term describes a rotor precessing at the Larmor frequency, while the off-resonant optical field provides the high-frequency Floquet modulation.

The resulting spin dynamics are formally analogous to those of a Kapitza pendulum as depicted in Fig.~\ref{fig:1}b. This correspondence enables the dynamical stabilization of the spin vector in its otherwise unstable orientations, allowing the atoms to exhibit a variety of unconventional magnetic orders with no equilibrium counterparts \cite{RN62}.
In the high-frequency limit ($\omega \gg \delta$), time-averaging the light-induced interactions via a Floquet-Magnus expansion generates nonlinear spin-transition terms that serve as a quantum analogue to the Kapitza potential \cite{SM}. This leads to a variant of the Lipkin-Meshkov-Glick (LMG) Hamiltonian,
\begin{equation}
\hat{H}_2 = \delta \hat{F}_{z} + \kappa \left( \hat{F}_{x}^{2} - \hat{F}_{y}^{2} \right),
\label{eq2}
\end{equation}
where $\hat{F}_i$ ($i=x,y,z$) denote the spin operators of the atom and the nonlinear coupling strength $\kappa = |\boldsymbol{E}|^2 \alpha_2 / 12$ is controlled by the optical intensity via the tensor polarizability $\alpha_2$.
This effective Hamiltonian exhibits a quantum phase transition at the critical point $\delta_c = 2F|\kappa|$, separating a paramagnetic phase ($|\delta| > \delta_c$) from a ferromagnetic phase ($|\delta| < \delta_c$). In the paramagnetic regime, the spin in the ground state is polarized along the $\hat{z}$ axis with no transverse spontaneous magnetization, and the sign of $\langle \hat{F}_z \rangle$ is determined by the sign of $\delta$. This transition is well described by mean-field theory \cite{RN57,RN58}.
The resulting critical dynamics map directly onto the inversion transition of the Kapitza pendulum and furnish the quantum-critical regime we exploit to enhance the metrological performance of the atomic sensor.

This connection becomes mathematically evident by mapping the spin dynamics onto a generalized harmonic oscillator via the Holstein-Primakoff transformation \cite{SM}. For an atom initially polarized along $\hat{z}$, we consider the short-time limit where the longitudinal component remains nearly saturated ($\hat{F}_z \approx F$). In this regime, the transverse spin dynamics are captured by the effective Hamiltonian
\begin{equation}
\hat H_3 = (1+g_{\mathrm{A}})\hat{p}^2 + (1-g_{\mathrm{A}})\hat{x}^2 + \eta\hat x,
\label{H3}
\end{equation}
where $\hat{x} = \hat{F}_x/\sqrt{F}$ and $\hat{p} = \hat{F}_y/\sqrt{F}$ are the canonical quadrature operators of the spin, and $g_{\mathrm{A}} = 2F\kappa/\delta$ is a dimensionless tuning parameter controlled by the ratio of optical intensity to the Zeeman field. A weak transverse magnetic field $B_x$, the target signal, introduces the linear perturbation $\eta \hat{x}$, where the effective coupling strength is $\eta = 2\sqrt{F}\gamma B_x/\delta$.

As illustrated in Fig. \ref{fig:1}b, for $g_{\mathrm{A}} < 1$ the concave effective potential corresponds to the dynamically stabilized regime of a Kapitza pendulum. The spin system then undergoes forced harmonic oscillation with normalized frequency $\Omega_{\mathrm{A}}  = 2\sqrt{1 - g_{\mathrm{A}}^2}$, effective mass $M_{\mathrm{A}} = 1/[2(1+g_{\mathrm{A}})]$, and the amplitude $A=\eta/(1-g_{\mathrm{A}})$. When $g_{\mathrm{A}} \approx 0$, the motion corresponds to small oscillations near the Bloch sphere's north pole. As $g_{\mathrm{A}}$ increases, both amplitude and period grow, reflecting an elliptization of the trajectory on the Bloch sphere.
For $g_{\mathrm{A}} > 1$, the normalized frequency $\Omega_{\mathrm{A}}$ becomes imaginary, indicating dynamical instability. Weak perturbations then grow exponentially, but the concurrent amplification of variance and the absence of a well-defined oscillation period preclude reliable sensing. This transition also marks a pseudo-anti-PT-symmetry breaking in the quantum system, a mechanism recently proposed to enable ultra-sensitive sensing in nonlinear wave-mixing platforms \cite{RN46}.
By operating at the edge of this ``dynamical equilibrium'', the sensor enters a regime of critical enhancement, where infinitesimal perturbations trigger a divergent response that is fundamentally unavailable in standard equilibrium-based architectures.
In the limit $g_{\mathrm{A}} \rightarrow 1$ (i.e., at criticality $\delta\rightarrow\delta_c$), the amplitude diverges, indicating strong signal enhancement even for weak perturbations. This is a direct manifestation of critical slowing down, where the oscillation period $\sim \Omega_{\mathrm{A}}^{-1}$ diverges and the harmonic potential softens, rendering the system increasingly responsive akin to a light free particle.

The critical amplification mechanism described above leads to a sensing protocol that is fundamentally distinct from squeezing-enhanced magnetometry, as illustrated in Fig. 1c.
In conventional squeezing protocols, the metrological gain relies on identifying and conditioning on specific correlations, such as between spins at different times or conjugate quadratures. Any loss or mixing of these correlations rapidly degrades the quantum enhancement.
In contrast, our critical protocol converts the quantum enhancement into a divergent susceptibility during the internal evolution, effectively disentangling the relevant degree of freedom before readout. The amplified signal can thus be extracted directly, without the need to reconstruct complex correlations, making the protocol intrinsically more robust than conditional squeezing schemes. This feature is conceptually analogous to the signal extraction in SU(1,1) interferometers \cite{RN101,RN102,RN85}, where the amplification is generated dynamically and does not require post-selection of correlated modes.

\begin{figure*}[tbph]
		\centering
		\includegraphics[width=\textwidth]{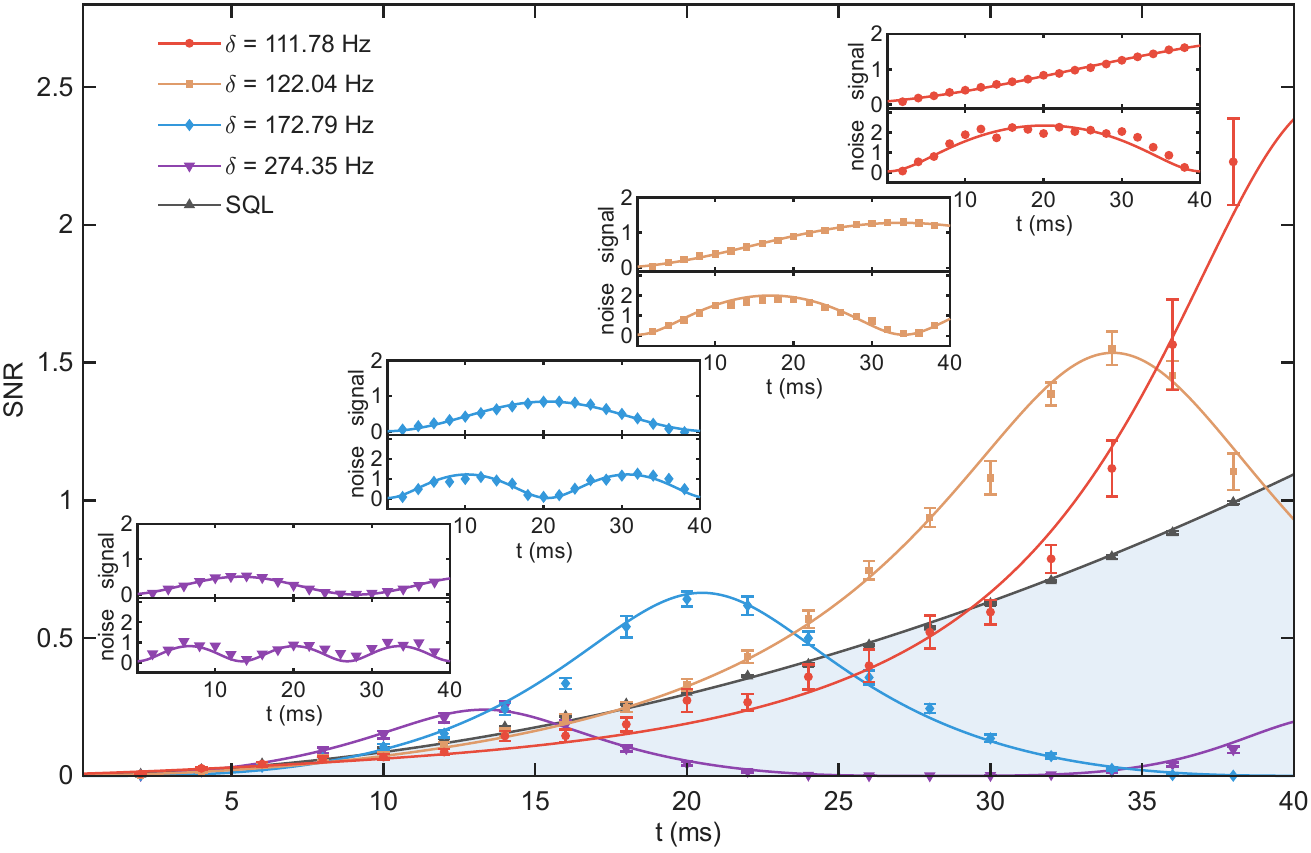}
        \caption{\justifying \textbf{Critical enhancement of magnetometer SNR}\\
SNR versus measurement time for the criticality-enhanced magnetometer at four detunings $\delta$ (symbols: inverted‑triangle, rhombus, square, circle), ordered from farthest to nearest the critical point. Solid curves are theoretical simulations. Triangles denote the conventional magnetometer, whose trace separates the performance regions above (pink) and below (blue) the SQL. Insets: Evolution of normalized signal $\langle\hat{F}_x(t)\rangle/\langle\hat{F}_y^{\mathrm{SQL}}(t=37.5\,\mathrm{ms})\rangle$ and normalized noise strength $10 \log_{10}[\Delta\hat{F}_x(t)/\Delta\hat{F}_x(0)]$. The noise oscillates with half the period of the signal, cycling from the coherent level to maximum amplification and back, thereby defining the optimal sensing instants.}
		\label{fig:2}
	\end{figure*}

\section{Periodically noiseless critical amplification toward supersensitivity}

A long-standing debate in critical quantum sensing centers on whether the divergent susceptibility at a critical point genuinely improves the SNR, given that quantum fluctuations are often amplified alongside the signal. We resolve this challenge by harnessing an engineered asymmetry between the dynamics of the signal mean and its variance near criticality. As defined by the relation $\mathrm{SNR} = (\partial_\eta \langle \hat{x} \rangle)^2 / \Delta^2 \hat{x}$, the metrological performance is determined by the competition between the squared dynamical susceptibility and the evolving noise floor.

As captured by the effective Hamiltonian (\ref{H3}), the squared susceptibility and the variance both diverge near criticality as $g_{\mathrm{A}} \to 1$, yet they exhibit a 2:1 periodicity contrast expressed as
\begin{eqnarray}
\left(\frac{\partial \langle \hat x(\tau) \rangle}{\partial \eta}\right)^2 & = & \frac{(\cos(\Omega_{\mathrm{A}}\tau)-1)^2}{4(1-g_{\mathrm{A}})^2}, \\
\Delta^2 \hat x(\tau) & = & \Delta^2 \hat x(0)\cos^2(\Omega_{\mathrm{A}}\tau) \nonumber\\
    &&+ (\frac{1+g_{\mathrm{A}}}{1-g_{\mathrm{A}}})\Delta^2 \hat p(0)\sin^2(\Omega_{\mathrm{A}}\tau),
\end{eqnarray}
This mismatch serves as the cornerstone of our noiseless critical amplification mechanism. The signal input, modeled by the linear interaction $\eta \hat{x}$, breaks the parity symmetry of the Hamiltonian and, as shown in Fig. 2(b), shifts the pendulum's equilibrium position away from $x=0$. Consequently, the initial state is no longer a steady state and undergoes periodic displacement, which results in a squared susceptibility that oscillates with a full period of $2\pi/\Omega_{\mathrm{A}}$. In contrast, the evolution of the variance $\Delta^2 \hat{x}$ is driven by the intrinsic quadrature mismatch between the initial state and the softened Hamiltonian. Because this variance evolution is a quadratic property of the wave function's shape rather than its mean displacement, it remains independent of the signal-induced shift and oscillates at twice the frequency.

This separation of scales creates a periodic ``noise-ebbing" window where the variance ebbs back to its initial coherent-state level even as the squared susceptibility reaches its divergent maximum at $\Omega_{\mathrm{A}}\tau = (2n+1)\pi$. By synchronizing the measurement with these optimal sensing durations, the system circumvents the parasitic noise inflation typically associated with critical gain. It is crucial to note that this measurement window depends on the specific form of the target field interaction. If the signal entered the Hamiltonian through higher-order or nonlinear terms that preserved the system's symmetries, such as those explored in previous theoretical frameworks \cite{RN15,RN46}, the periodicity contrast could vanish, precluding the possibility of noiseless amplification. Within our framework, however, the synchrony between peak critical gain and periodic noise restoration maximizes the performance to $\mathrm{SNR}_{\mathrm{max}} = [ (1-g_{\mathrm{A}})^2 \Delta^2 x(0) ]^{-1}$, effectively surpassing the standard quantum limit using only easily prepared spin coherent states.

To benchmark this enhancement, we apply a weak transverse magnetic field $B_{x} = 0.14\,\text{nT}$ and measure both the mean value $\langle\hat{F}_x(t)\rangle$ and the fluctuation $\Delta\hat{F}_x(t)$ of the transverse spin component $\hat{F}_x(t)$ while tuning the Zeeman detuning $\delta$ toward the critical value $\delta_c=78.58\,\text{Hz}$ under a fixed nonlinear coupling $\kappa$.
The spin component $\hat{F}_x$ is monitored through a quantum non-demolition (QND) interface utilizing a far-detuned, linearly polarized probe beam along the $x$-axis. The interaction is described by $\hat{H}_{\mathrm{QND}}=\alpha\hat{S}_x\hat{F}_x$, effectively mapping the spin fluctuations onto the optical Stokes operator $\hat{S}_y$ \cite{RN9}, which is then resolved by a polarimeter.

A universal challenge in critical sensing is that critical slowing down confines measurements to a low-frequency regime where the fundamental spin-projection noise is often overwhelmed by technical noises (e.g., magnetic-field drifts and laser instabilities). To recover these noises and thus access the true quantum-limited sensitivity, we employ a differential measurement scheme. In the experiment, the atomic ensemble evolves subject to the critical dynamics and the signal field $B_x$. The measurement sequence is repeated over a duration of $10\,\text{s}$ to experimentally acquire 250 independent estimates of $\hat{F}_{x}$. Slow drifts are rejected by subtracting the initial value of each sequence, akin to lock-in detection. The set of 250 differences is then Fourier analyzed to produce a noise spectrum.
Residual non‑spin fluctuations are further suppressed by a Kalman filter based on the known critical spin dynamics.
Finally, the resulting spectrum aligns with calibrated expectations, verifying the quantum nature of the noise.

	
To quantify the metrological gain, $G=\Delta^2 B_x^{\mathrm{SQL}}/\Delta^2 B_x$, we compare our sensor’s performance with the SQL.  This limit corresponds to the optimal sensitivity of a conventional magnetometer ($\delta = \kappa = 0$), in which spins precess in the $\hat{y}$-$\hat{z}$ plane under $B_x$, and the linear growth of $\langle \hat{F}_y(t) \rangle$ over short times provides $\mathrm{SNR_{SQL}}=2Ft^2$. Throughout this work, all signals, noise levels, and SNRs are normalised to the values obtained with the conventional magnetometer at the fixed working time $t=37.5$ ms. Experimentally we observe a maximum $\mathrm{SNR}$ improvement factor of $2.13$, equivalent to a metrological gain of $3.3$ dB.

The main panel of Fig. \ref{fig:2} displays the measured SNR versus evolution time for various detunings $\delta$, corresponding to decreasing distance from the critical point $\delta_c$. Insets illustrate that signal and noise oscillate periodically, the noise period being half that of the signal in every case. As $\delta$ approaches $\delta_c$, the oscillations slow down and their amplitude grows, reflecting both critical slowing down and divergent susceptibility. The SNR peaks at the normalized time $\tau_{p}=\pi/2\sqrt{1-g_{\mathrm{A}}^2}$, which coincides with the first complete period of the noise and half the period of the signal. In the ideal coherent limit this peak time would diverge at the critical point; in our experiment the finite coherence time ($45.86\, \text{ms}$) sets the maximum usable evolution duration and hence the attainable $\mathrm{SNR}_{\mathrm{max}}$.


\begin{figure*}[tbph]
		\centering
		\includegraphics[width=\textwidth]{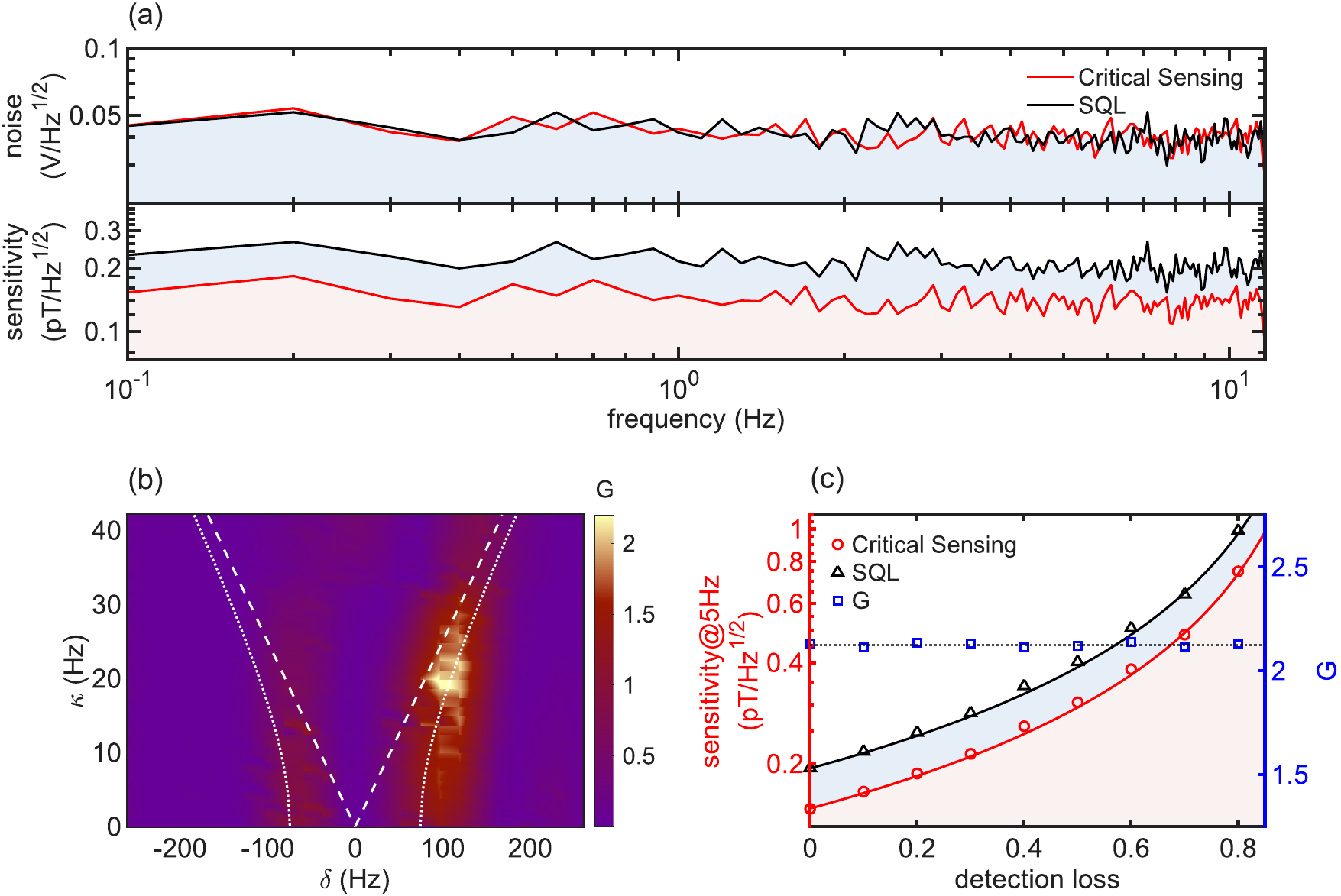}

        \caption{\justifying \textbf{Metrological gain and loss-robustness of criticality-enhanced sensing}\\
(a) Noise level and magnetic-field sensitivity with (red) and without (black) criticality enhancement. The SQL for $B_x$ measurement is $200.54\, \rm fT/\sqrt{Hz}$; at the optimal working point the criticality-enhanced sensitivity reaches $137.69\, \rm fT/\sqrt{Hz}$. (b) Metrological gain $G$ versus nonlinear coupling strength $\kappa$ and detuning $\delta$ at a fixed evolution time $t = 37.5$ ms. White dashed lines mark the critical condition $\kappa = |\delta|/(2F)$, and white dotted line mark the working points $\delta^2-(2F\kappa)^2=(\pi/t)^2$. (c) \textit{Left vertical scale}: Sensitivity (circles) under increasing detection loss for critical sensing, compared with the SQL (triangles). \textit{Right
vertical scale}: Corresponding metrological gain, which remains robust against detection loss.}

		\label{fig:5}
	\end{figure*}

From $\mathrm{SNR}_{\mathrm{max}}$ and the relation $\eta = 2\sqrt{F}\gamma B_x/\delta$ we derive the magnetic-field sensitivity for an initial spin-coherent state $(\Delta^{2}x(0)=\Delta^{2}p(0)=1/2)$ \cite{SM}. Ignoring the short preparation time $\tau_{1}$ and readout time $\tau_{3}$ (Fig. \ref{fig:1}c), the sensitivity over the coherent sensing interval $\tau_{2}$ is
\begin{equation}
    \Delta B_{x} = \frac{\delta }{2^3\sqrt{2F}(1+g_{\mathrm{A}})\gamma (\tau_{2}/\pi)^{2}},
\end{equation}
where $F$ is the total spin length. Notably, this sensitivity saturates the quantum Cramer-Rao bound, $\Delta^{2}B_{x} \geq 1/\mathcal{F}_{B_{x}}$, where $\mathcal{F}_{B_{x}}$ is the quantum Fisher information associated with the initial coherent state $|\psi_0\rangle$ and the evolution Hamiltonian $H_3$. This equivalence confirms that our chosen observable $\hat{x}$ (the collective spin component $\hat{F}_x$) serves as an optimal estimator, fully extracting the Fisher information encoded within the critical quantum dynamics.
The resulting scaling $\Delta^{2}B_{x} \propto 1/(F\tau_2^4) $ surpassing the SQL of conventional sensors $(\Delta^{2}B_{x} \propto 1/(F\tau_{2}^{2}))$ by a factor $\tau_{2}^{-2}$. The improvement originates directly from the critical divergence of the susceptibility $\partial_\eta\langle\hat x\rangle$.
In comparison, squeezing-enhanced sensors achieve $\Delta^2 B_x \propto 1/(\mathcal{S} F \tau_2^2)$, where $\mathcal{S}$ is the squeezing degree. However, as Fig. \ref{fig:1}(c) shows, a trade-off between increasing $\mathcal{S}$ and reducing $\tau_2$ constrains the overall metrological gain.

To quantify the sensitivity experimentally, we fix the evolution time at $t=37.5$ ms and acquire the noise spectrum using the differential measurement and Kalman‑filtering procedure described earlier. The upper panel of Fig. \ref{fig:5}a shows that the noise of the critical sensor matches that of the conventional sensor, confirming that the spin-projection noise at measurement instants remains at the coherent-state level. While the noise is unchanged, the sharply enhanced susceptibility near the critical point translates directly into improved sensitivity (lower panel of Fig. \ref{fig:5}a), reaching an optimum of $137.69\,\text{fT}/\sqrt{\text{Hz}}$.

Importantly, unlike squeezing-enhanced protocols, whose quantum advantage relies on fragile spin correlations that are highly sensitive to detection efficiency, this critical enhancement directly converts the quantum advantage into a divergent susceptibility. This mechanism renders the metrological gain inherently robust to detection loss. As shown in Fig. \ref{fig:5}c, the gain remains nearly constant as detection loss increases, and the sensitivity of the critical sensor consistently exceeds that of the conventional sensor across the full loss range. This behaviour confirms the intrinsic resilience of the critical-sensing mechanism to non-ideal detection.

Because the critical enhancement depends on how closely the system approaches the critical point, precise control and maintenance of $\delta$ and $\kappa$ during sensing are essential. The Zeeman detuning frequency $\delta$ can be controlled with high accuracy through a precision current source (Yokogawa, GS200) driving the Helmholtz coils, and the nonlinear coupling $\kappa$, proportional to the optical intensity, has been carefully calibrated (see Methods). We therefore measure the metrological gain $G$ across the $\kappa$-$\delta$ parameter space (Fig. \ref{fig:5}b).

The critical boundaries are defined by the condition $\kappa = |\delta| / (2F)$. The optimal evolution time $t_p$, at which the spin noise returns to its initial level, depends on both $\kappa$ and $\delta$ and follows the hyperbolic relation $\delta^2 - (2F\kappa)^2 = (\pi/t_p)^2$, diverging as the system approaches criticality. Gain emerges in the region near the boundary for $\delta > 0$ (the sign of $\delta$ would flip if the initial polarization of $\hat{F}_z$ were reversed). A maximum metrological gain of $3.3$ dB is achieved at $\kappa = 19.65$ Hz and $\delta = 114.86$ Hz, lying on the hyperbolic trajectory consistent with the optimal $t_p$ identified in the SNR analysis. The subsequent decrease in gain for larger $\kappa$ is attributed to light-shift-induced decoherence, which shortens the coherence time and limits the usable evolution duration.

\section{Conclusion and outlook}

In summary, our work demonstrates a universal protocol for quantum-enhanced metrology that exploits the non‑equilibrium dynamics near a dynamical critical point. By realizing this principle in a light-driven atomic spin ensemble, we resolve the long-standing debate regarding the metrological utility of quantum criticality. We show that the fundamental asymmetry between signal and noise evolution near a stability threshold, specifically the periodic ``ebbing" of quantum fluctuations, allows a sensor to act as a built-in noiseless amplifier. This mechanism enables the system to capture a nonlinearly divergent signal while the noise floor periodically returns to its coherent baseline, achieving a magnetic sensitivity of $137.69\text{ fT}/\sqrt{\text{Hz}}$ and a $3.3\text{ dB}$ metrological gain over the SQL.

This criticality-enhanced approach addresses a key bottleneck in conventional quantum metrology. Unlike spin-squeezing or many-body entanglement, which offer high theoretical gains but are notoriously fragile to technical noise and state-preparation imperfections, our protocol derives its advantage from the system's intrinsic dynamical scaling. This makes the enhancement inherently more robust against the “quantum fragility” that often prevents non-classical states from reaching their theoretical potential in practical environments. While the absolute gain in our current room-temperature setup is limited by the atomic coherence time, a constraint shared by all atomic sensors, this is a technical rather than a conceptual limitation. Transitioning to platforms with longer coherence, such as thermal atomic ensembles with coherence time on the order of seconds \cite{RN67}, ultra-cold atoms \cite{RN80,RN81} or hybrid alkali-noble-gas systems \cite{RN82,RN83,Rn84}, will allow the system to evolve closer to the critical point for longer durations, fully unlocking the divergent scaling of the SNR.

Importantly, this framework is a complementary architecture rather than a mere substitute for non-classical resources. While our current demonstration uses a coherent initial state to highlight the noiseless amplification of the critical dynamics themselves, future hybrid schemes could integrate spin-squeezed states to amplify a pre-existing non-classical advantage. Such a combination could push sensitivities into the sub-femtotesla regime. This versatility ensures that criticality-enhanced sensors can be tailored to diverse demands, ranging from field-deployable biomagnetic imaging where operational robustness is paramount, to fundamental physics searches such as the detection of ultralight dark matter where the quest for the sensitivity limit justifies the technical overhead of non-classical state preparation.

  \section{Methods}

    \subsection{Quantum Fisher Information and SNR}

  To leverage the property for highly sensitive measurement of the transverse magnetic field $B_x$, we calculate the Quantum Fisher Information (QFI) for $\eta$. QFI determines the maximum amount of information that can be extracted for a certain parameter from the system, and can be written for a pure probe state $\left|  \psi_0\right\rangle$ with unitary process as \cite{RN12}:
	\begin{equation}
		\mathcal{F}_a=4\left( \left\langle \psi_0\right| \hat{\mathcal{H}}^2_a\left|  \psi_0\right\rangle -\left| \left\langle \psi_0\right| \hat{\mathcal{H}}_a\left|  \psi_0\right\rangle\right| ^2\right)
		\label{eq9}
	\end{equation}
	where $\hat{\mathcal{H}}_a=-i \mathrm{e}^{i\hat{H}t}\partial_a \mathrm{e}^{-i\hat{H}t}$.
            	
\begin{figure}[h]
		\centering
		\includegraphics[width=0.45\textwidth]{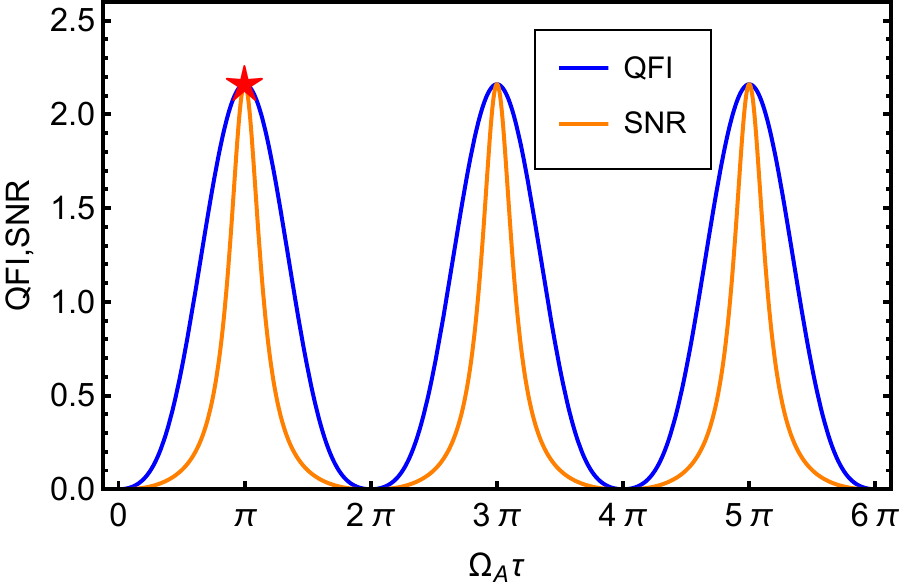}
		\caption{\justifying The blue line is the QFI for detecting $\eta$, the yellow line is the SNR of measuring $\langle\hat{F}_x\rangle$, and the black line is the SQL for detecting $\eta$. The red star is the first working time in which the SNR oscillates to the maximum value and reach the QFI. All data are normalized to the SQL at the the first working time.}
		\label{sfig:1}
        \end{figure}

    Also, we calculate the SNR and give the explicit expression in the Holstein-Primakoff transformation as:
    \begin{widetext}
	\begin{equation}
		\begin{aligned}			\mathrm{SNR}&=\dfrac{\partial_{\eta}\langle \hat{x}_{\theta}(\tau)\rangle ^2}{\Delta^2\hat{x}_{\theta}(\tau)}\\
			&=\dfrac{[-2 \Omega_{\mathrm{A}} \mathrm{cos} (\theta) \mathrm{sin}^2 \left(\frac{1}{2}\Omega_{\mathrm{A}} \tau\right)+(1-g_{\mathrm{A}}) \mathrm{sin} (\theta) \mathrm{sin} ( \Omega_{\mathrm{A}}\tau)]^2}{2(1-g_{\mathrm{A}})^2\{1+g_{\mathrm{A}}[-g_{\mathrm{A}} \cos(2  \Omega_{\mathrm{A}}\tau)+2 \cos (2 \theta) \sin ( \Omega_{\mathrm{A}}\tau)^2-\Omega_{\mathrm{A}} \sin (2 \theta) \sin (2 \Omega_{\mathrm{A}}\tau )]\}}
		\end{aligned}
	\end{equation}
    \end{widetext}
    where $\hat{x}_{\theta} $ denotes the transverse spin component in the direction of $\theta$ relative to the $x$ axis.  
    For $\theta=0$, we have $\hat{x}_{\theta}=\hat{x}$, cooresponding to the spin component along the $x$ axis. By comparing the SNR for this observalbe with QFI, we find that both quantities oscillate periodically in time and reach their maxima simultaneously at the working time, as shown in Fig. \ref{sfig:1}.
The simultaneous maximization of SNR and QFI confirms that $\hat{x}$ serves as an optimal estimator, extracting the full quantum information encoded by the critical dynamics.

\subsection{Experimental platform}
	\begin{figure}[h]
		\centering
\captionsetup{justification=raggedright,singlelinecheck=false}
			\centering
			\includegraphics[width=0.5\textwidth]{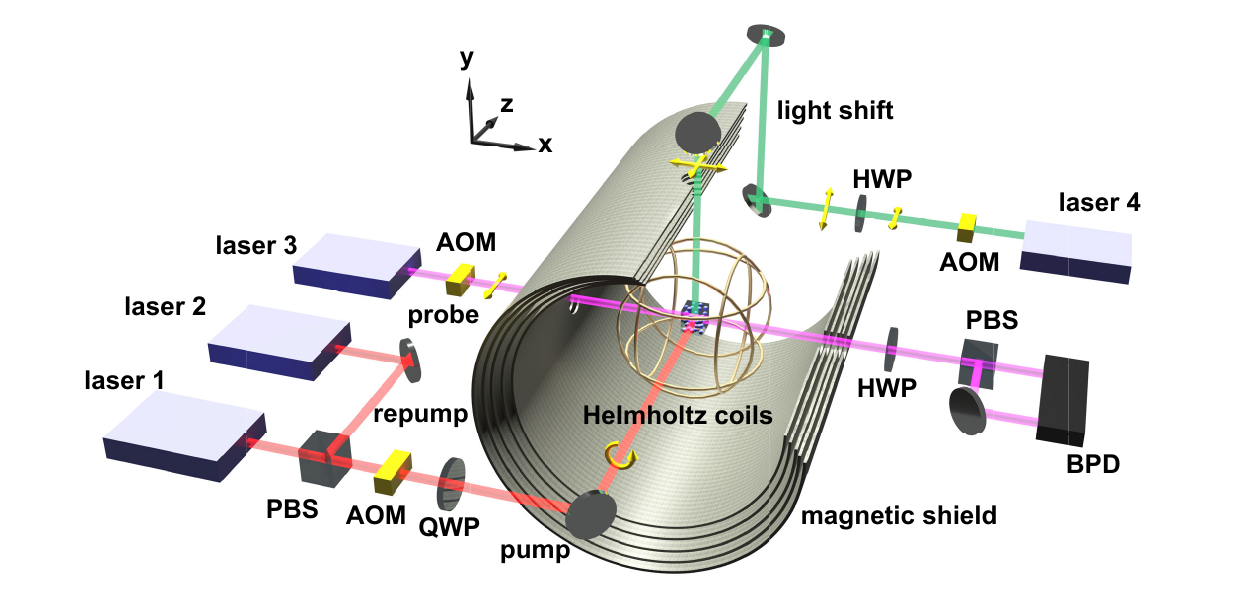}
		\vfill		
		\caption{\justifying The experimental schematics. PBS: polarization beam splitter, QWP: quarter-wave plate, AOM: acousto-optical modulator, HWP: half-wave plate, BPD: balanced photodetector.}
		\label{sfig:2}
	\end{figure}
	The experimental setup is illustrated in Fig. \ref{sfig:2}. We work with a room temperature ensemble of $1.35\times 10^{11}$ $^{87}\mathrm{Rb}$ atoms contained in a cubic glass cell with a side length of 3 cm. The cell is located at the center of a multi-layer magnetic shield with 3D Helmholtz coils designed to control the magnetic field in the $x$, $y$ and $z$ directions. Two 795 nm lasers are used to produce pump and repump light beam sharing the same circular polarization to excite the atomic transition of $\left| 5S_{1/2},F=2\right\rangle \to \left| 5P_{1/2},F=2\right\rangle $ and $\left| 5S_{1/2},F=1\right\rangle \to \left| 5P_{1/2},F=2\right\rangle $ separately. Another 795 nm laser is far-detuned from the optical resonance to induce tensor light shift.
    The polarization direction of this off-resonant light is adjusted by a half-wave plate to exactly $35.26^{\degree}$ with respect to  $x$-axis. 
    Three acousto-optic modulators (AOMs) are employed to control the switching of the light pulses in the three-stage sequence illustrated in Fig. \ref{fig:1}. In the first stage, the pump laser (2.4 mW) and repump laser (8.2 mW) are applied simultaneously for $\tau_1=2 $ ms to optically pump the atoms into the initial state $|F=2, m_{F}=2\rangle$. Following state preparation, the spins evolve under LMG Hamiltonian. Here, the linear term $\delta \hat{F}_{z}$ is controlled by the bias magnetic field applied along the $z$-axis, while the nonlinear coupling $\kappa$ is induced by an off-resonant laser beam. The strength of $\kappa$ can be tuned by adjusting the intensity and frequency of this beam. After an evolution time corresponding to the optimal working point, a $391\,\mu$W, far-detuned, and $z$-linearly polarized 780 nm probe laser is applied to perform a quantum non-demolition (QND) measurement via the interaction $\hat{S}_y \propto\ \hat{F}_x$ \cite{RN9}, where $\hat{S}_y$ is the Stokes operator of probe light. The signal $\hat{S}_y$ is read out using a combination of a polarization beam splitter and a balanced photodetector, allowing us to monitor the evolution of $\hat{F}_x(t)$.
    All lasers are actively stabilized in both power and frequency to suppress technical noise. Power stabilization is implemented using noise eaters for the pump, repump, probe and the off-resonant lasers. For frequency stabilization, the pump and repump lasers are locked directly to the atomic resonance via saturated absorption spectroscopy. For the off-resonant light used to generate the tensor light shift and the probe light used for QND measurement, an electro-optic modulator (EOM) is employed to generate frequency sidebands. The first-order sideband is locked to the atomic resonance, such that the carrier frequency is detuned by an amount precisely determined by the EOM modulation frequency.

   \subsection{calibration of $\kappa$}
     \begin{figure}[h]
		\centering
		\includegraphics[width=0.5\textwidth]{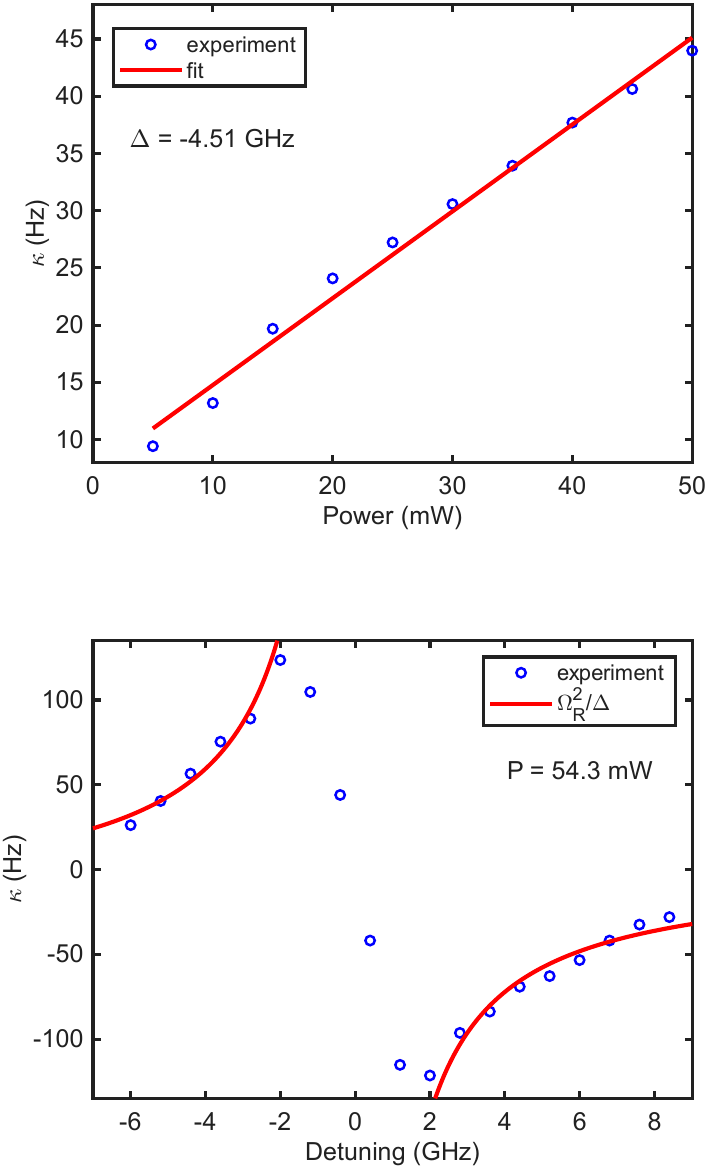}
		\caption{\justifying The AC-Stark shift for different light power and detuning. The shift in far-detuned region are fitted to $\Omega_R^2/\Delta$ with $\Omega_R$ is the Rabi frequency of light.}
		\label{sfig:3}
	\end{figure}
    The nonlinear term in LMG Hamiltonian in our experiment is generated by the tensor light shift. To calibrate the nonlinear coupling strength $\kappa$, we use the fact that a $z$-polarized off-resonant light field produces an effective Hamiltonian of the form $3\kappa\hat{F}_{z}^2$, which induces a splitting in the magnetic resonance. By measuring this splitting, we calibrate the nonlinear coupling strength. The measured splittings change with power and detunning are shown in Fig. \ref{sfig:3}.

    \subsection{Measurement of critical tuning parameter}
       \begin{figure}[tbph]
		\centering
	\includegraphics[width=0.5\textwidth]{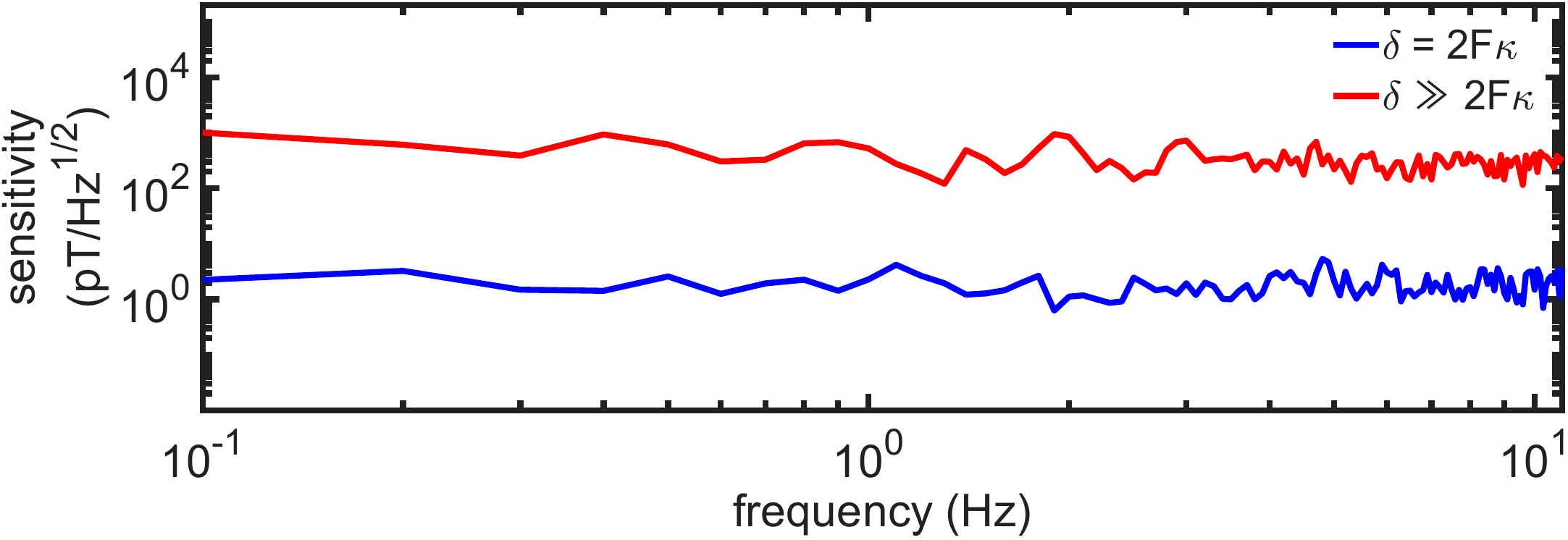}
		\caption{\justifying Magnetic-field sensitivity at the critical point (Blue) and far away from the critical point (Red) for $B_z$ measurement.}
		\label{sfig:9}
	\end{figure}
  It is worth noting that, due to the typical critical behavior of quantum systems with $SU(1,1)$ symmetry, an even better sensitivity scaling of $1/\tau^6$ could be achieved if the sensed parameter were the critical tuning parameter $g$ (which is determined by $B_{z}$ and $\kappa$) \cite{RN63,RN64,RN46} rather than the external perturbation strength $\eta(B_x)$ considered here.
  As shown in Fig. \ref{sfig:9}, the sensitivity of $B_z$ at the critical point gives an metrological gain of 41.84 dB.
  However, in this measurement scheme, the effective number of atoms that participate in the sensing process is given by the number of bosonic excitations \cite{RN46}, which is required to be much smaller than the total atom number \cite{RN8,RN58}.
 In contrast, our sensing scheme harnesses all atoms for the measurement, and the standard quantum limit of conventional magnetometer we compare against is defined with respect to the total atom number. Therefore, while measuring $B_{z}$ offers a better metrological gain, its absolute sensitivity is in inferior to that achieved by $B_{x}$ measurement.


	\bibliography{reference}

	\begin{acknowledgments}
		We thank Guiying Zhang from Zhejiang University of Technology for providing the paraffin-coated atomic cell. This work was supported by the Innovation Program for Quantum Science and Technology (2021ZD0303200 and 2024ZD0302200), the National Natural Science Foundation of China (grant nos. 12234014, 12374328, 12204303 and 11654005), the Shanghai Municipal Science and Technology Major Project (2019SHZDZX01), Shanghai Science and Technology Innovation Action Plan (24LZ1401400 and 24LZ1400600), and the National Key Research and Development Program of China under grant number 2016YFA0302001. W.Z. also acknowledges additional support from the Shanghai Talent Program.
	\end{acknowledgments}

    \section*{Author Contributions}
Supervision: W.Z. Conceptualization: G.B., K.Z., and W.Z. Investigation: D.H., M.S., G.B., K.Z., and W.Z. Methodology: G.B., K.Z., and W.Z. Data curation: D.H., and G.B. Formal analysis: D.H., G.B., K.Z., and W.Z. Visualization: D.H., G.B., and K.Z. Writing—original draft: D.H., and G.B. Writing—review and editing: G.B., K.Z., and W.Z.
\end{document}